%% file: main.tex
\documentclass[times,preprint,11pt,nopreprintline]{elsarticle}
\ifdefined\pdfminorversion\pdfminorversion=7\fi
\usepackage[T1]{fontenc}
\usepackage{tgtermes}
\biboptions{numbers,sort&compress}
\usepackage{amsmath,amssymb,amsfonts}
\usepackage{graphicx,textcomp}
\usepackage{multirow,booktabs,array}
\usepackage{xcolor,placeins,geometry}
\usepackage{hyperref,xurl}
\hypersetup{hidelinks}
\input{arxiv_format}

\begin{document}
\hypersetup{pdfauthor={Jiahao Qin; Yiwen Wang},pdftitle={M\"obius Interaction Filtering: Order-Selective Spectral Regularization for Multimodal Humor Detection}}
\begin{frontmatter}
\title{M\"obius Interaction Filtering: Order-Selective Spectral Regularization for Multimodal Humor Detection}
\author{Jiahao Qin\corref{cor1}}
\ead{jiahao.qin19@gmail.com}
\author{Yiwen Wang}
\cortext[cor1]{Corresponding author.}

\begin{abstract}
Multimodal humor detection requires interpreting language, vocal delivery, and visual behavior jointly. Expressive fusion must also regulate how interactions of different orders enter the final judgment. We propose M\"obius Interaction Filtering (MoIF), which learns a modality-subset prediction function and imposes order-selective spectral regularization through its aggregation structure. Private temporal encoders and balanced reciprocal interaction construct supervised unimodal, bimodal, and trimodal predictors. Their uniform aggregate is exactly a M\"obius-domain filter with a geometrically decreasing interaction-order response. This formulation retains higher-order representations while regulating their contribution to prediction. Controlled UR-FUNNY perturbations reveal a pronounced decline in label-aligned signal-to-noise ratio across interaction orders. The induced filter improves the final ratio by 1.62--2.79 times relative to the co-trained trimodal predictor while preserving its own clean class-aligned signal under perturbation. Five-seed experiments achieve 70.6\% accuracy and weighted F1 on UR-FUNNY, exceeding thirteen reproduced methods on these detection metrics. CMU-MOSI and CMU-MOSEI evaluations establish applicability to sentiment prediction, and cross-seed analyses show more stable estimates on all three datasets. MoIF connects the organization of multimodal interactions to their statistical reliability through an executable spectral constraint.
Code is available at \url{https://github.com/D-ST-Sword/MoIF}.
\end{abstract}
\end{frontmatter}

\section{Introduction}
\label{sec:intro}
Humor is expressed through the interplay of words, vocal delivery, and visual behavior. A verbal statement may carry an apparent meaning that is reinforced, qualified, or overturned by its delivery. Multimodal humor detection therefore requires learning relationships between these streams while retaining evidence that is already informative within an individual modality \cite{hasan2019urfunny,poria2017review,baltrusaitis2019multimodal}. The challenge is not simply to construct increasingly comprehensive combinations. It is to determine how evidence from different interaction orders should contribute to a stable final judgment.

Existing fusion methods offer increasingly expressive representations. Tensor fusion enumerates unimodal and cross-modal feature combinations, and low-rank factorization reduces their computational cost \cite{zadeh2017tensor,liu2018efficient}. Cross-modal attention relates unaligned temporal streams \cite{tsai2019multimodal}, while shared--private learning and unimodal supervision preserve modality-specific information \cite{hazarika2020misa,yu2021learning}. Dynamic Fusion Graph and Multimodal Routing explicitly organize information from different interaction orders \cite{zadeh2018mosei,tsai2020routing}. Recent interaction-aware models further specialize representations or experts for complementary information \cite{xin2025i2moe,yang2026dmil}. These approaches develop interaction representations and learned rules for their combination. We focus on the structure of the prediction operator: how its response across interaction orders relates expressiveness to estimation stability.

Why not rely exclusively on the representation that combines all modalities? At the information level, complete inputs contain every subset. At the estimator level, however, a learned trimodal predictor is not guaranteed to dominate separately supervised subset predictors: finite data, shared parameters, and joint optimization determine which relationships are captured and how reliably they are estimated. Multimodal training studies document uneven learning across streams and insufficient exploitation of useful unimodal features \cite{wang2020training,du2023unimodal,peng2022balanced}. Retaining unimodal and bimodal predictions gives the model access to several estimates of the same target, but a useful design must also explain what their combination does to the interaction structure.

We address this question through M\"obius Interaction Filtering (MoIF). The central design variable is not only which modalities are fused, but how the resulting interaction spectrum enters the decision. Predictions indexed by modality subsets define a set function whose M\"obius representation resolves increments associated with individual modalities, pairs, and the complete triplet \cite{grabisch2000equivalent}. Complete-input evaluation sums these increments with unit response at every order. MoIF instead induces a geometrically decreasing response through uniform subset aggregation. This connects the prediction rule to discrete-set signal processing \cite{puschel2021discrete}: the model retains expressive interactions while its aggregation structure regulates their contribution to the decision.

MoIF realizes this principle as a learned lifting into the modality-subset function space, followed by a fixed prediction operator. Private temporal encoders, balanced reciprocal interaction, and trimodal composition jointly map an observation to supervised predictions across the subset lattice. Direct subset supervision retains task-relevant estimates at each input combination, while the primary objective trains their filtered decision. The learned function and the spectral constraint are thus coupled throughout optimization. Subset aggregation implements the operator exactly, without an explicit transform or a separate filtering network.

The learned interaction spectrum has a measurable reliability structure. On UR-FUNNY, controlled feature perturbations reveal descending label-aligned signal-to-noise ratios from first- to third-order components across every tested seed--condition combination. MoIF's geometric response attenuates the less favorable orders without discarding them, and the combined estimate achieves a higher signal-to-noise ratio while retaining its class-aligned response. Figure~\ref{fig:spectral_mechanism} connects the coordinate transformation, filter response, and these measured effects. Humor detection is the primary task; separately trained CMU-MOSI and CMU-MOSEI evaluations test the same interaction principle on sentiment prediction within the broader setting of multimodal affective computing. Cross-seed stability provides a complementary test of the resulting estimates.

Our contributions are threefold. First, we formulate multimodal prediction as the design of an interaction-spectrum response and derive the exact subset-domain implementation of an order-selective M\"obius filter. This separates the construction of expressive interactions from their weighting in the final estimate. Second, we develop a jointly supervised lifting with balanced reciprocal interaction, coupling private, pairwise, and trimodal predictors to the filtered task objective. Third, we connect the measured decline in interaction-order reliability to improved signal-to-noise balance under the induced response, supported by controlled perturbations and five-seed humor-detection and sentiment evaluations. Together, these contributions establish an executable spectral constraint linking multimodal expressiveness to estimation stability.

\section{Related Work}
\label{sec:related}
\subsection{Multimodal Interaction and Humor Detection}
UR-FUNNY studies humor through language, acoustic cues, and visual behavior \cite{hasan2019urfunny}. Its multimodal formulation connects humor detection to broader work on sentiment and affective prediction \cite{poria2017review,baltrusaitis2019multimodal}. TFN and LMF model feature combinations explicitly \cite{zadeh2017tensor,liu2018efficient}; MulT uses directional attention to capture dependencies between unaligned streams \cite{tsai2019multimodal,vaswani2017attention}. TETFN retains unimodal predictions alongside text-oriented interactions \cite{yang2023tetfn}, while ALMT and DLF investigate language-guided representation learning and disentangled fusion \cite{zhang2023almt,zeng2025dlf}. BC-PMJRS combines mutual-information objectives in a predefined joint representation space \cite{qin2025bcpmjrs}. TLRRB develops text-anchored token-pair regions with supervised contrastive refinement and cross-modal alignment \cite{li2026tlrrb}. MoIF uses balanced reciprocal interaction to construct subset predictors, then organizes their decision contributions through the M\"obius interaction filter.

\subsection{Specialization, Collaboration, and Input Quality}
MISA separates shared and private representations, and Self-MM supplies modality-specific learning signals \cite{hazarika2020misa,yu2021learning}. Unequal learning dynamics and conditional modality utilization motivate more deliberate organization of multimodal optimization \cite{wang2020training,du2023unimodal,wu2022greedy,peng2022balanced}. GCE improves complementary representation learning through a grading-inspired design \cite{huang2026grading}. Structured methods address collaboration directly: DFG connects unimodal, bimodal, and trimodal representations; Multimodal Routing learns their contribution to output concepts; I$^2$MoE specializes experts for interaction types; and DMIL learns an information-theoretic interaction decomposition \cite{zadeh2018mosei,tsai2020routing,xin2025i2moe,yang2026dmil}. MoIF directly supervises subset predictions and derives a fixed interaction-order response from their aggregation. Its distinction is the explicit connection between that response and measured interaction reliability, beyond organizing multiple pathways or learning their routing weights.

Input degradation provides another motivation for structured fusion \cite{wu2026deepmissing,zheng2024niat}. EMT restores multimodal features at two levels, while MPLMM, LNLN, and CASP study learning and adaptation with unavailable modalities \cite{sun2023emtdlfr,guo2024mplmm,zhang2024lnln,li2025casp}. MoIF addresses the reliability of interaction contributions within a jointly learned prediction function. Its explicit subset dependencies also make it possible to examine how a modality changes the eligible interaction structure and the resulting estimate.

\subsection{Set Functions and Interaction Spectra}
A set function admits equivalent representations in subset-value and M\"obius coordinates \cite{grabisch2000equivalent}. Discrete-set signal processing associates such representations with filtering operators and spectra indexed by subsets, with cardinality providing an interaction-order interpretation \cite{puschel2021discrete}. We apply this mathematical structure to a family of jointly learned modality-subset predictions. The resulting filter has a closed-form response determined by subset inclusion. This response defines MoIF's implicit order-selective spectral regularization on the interaction components of its learned prediction function.

\section{M\"obius Interaction Filtering}
\label{sec:method}
\begin{figure*}[p]
\centering
\includegraphics[width=\linewidth,height=0.80\textheight,keepaspectratio]{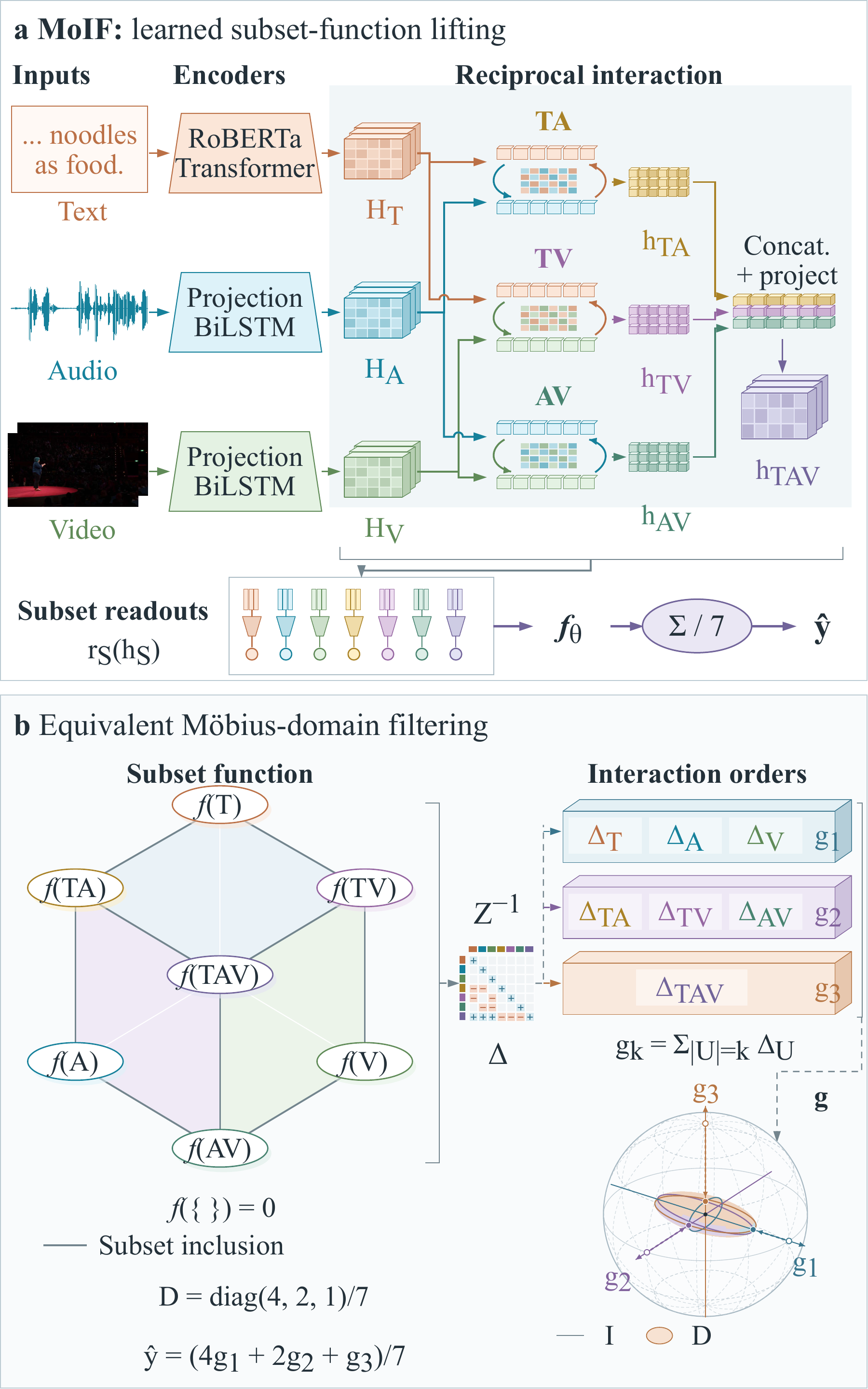}
\caption{MoIF architecture and equivalent order-selective filtering. (a) Private encoding, reciprocal interaction, and trimodal composition learn seven subset readouts; their mean yields the prediction. (b) Subset values map to M\"obius increments and order sums. In order-sum coordinates, the gray unit sphere and colored ellipsoid show the responses of $I$ and $D=\operatorname{diag}(4,2,1)/7$. Inward arrows connect unit-axis responses $e_k$ to $De_k$. Coordinate summation gives $f(TAV)$ for $I$ and MoIF for $D$. Dashed links indicate equivalent coordinate views. Tensor tiles are schematic; input media are from UR-FUNNY.}
\label{fig:overview}
\end{figure*}

MoIF learns a lifting from multimodal observations into a modality-subset function space, then applies an order-selective prediction operator. The lifting constructs task-directed evidence across input combinations; the operator regulates its contribution in M\"obius coordinates. We define this composition, derive its spectral response, and specify the neural parameterization and joint objective. Figure~\ref{fig:overview} connects these elements in one model.

\subsection{Learned Lifting into the Modality-Subset Function Space}
Let $\mathcal M=\{T,A,V\}$ index text, audio, and vision, with $M=|\mathcal M|=3$. Modality $m$ provides a sequence $X_m\in\mathbb R^{L_m\times d_m}$ and a validity mask $b_m\in\{0,1\}^{L_m}$. We write $x_S=\{(X_m,b_m):m\in S\}$ for the inputs belonging to subset $S$. Humor labels are $y\in\{-1,+1\}$, and the final score is thresholded at zero; continuous sentiment targets use the same scalar prediction formulation.

The subset lattice $(2^{\mathcal M},\subseteq)$ organizes the nested input combinations. Let $\mathcal V_{\mathcal M}=\{f:2^{\mathcal M}\to\mathbb R\mid f(\varnothing)=0\}$ be its real-valued function space, with $2^M-1$ coordinates. MoIF learns a lifting $\Phi_\theta:\mathcal X\to\mathcal V_{\mathcal M}$ from the observation space $\mathcal X$:
\begin{equation}
\begin{aligned}
\bigl[\Phi_\theta(x)\bigr](S)&=f_\theta(S;x)=r_S\!\left(h_S(x_S;\theta)\right),\\
&\varnothing\ne S\subseteq\mathcal M,\qquad [\Phi_\theta(x)](\varnothing)=0.
\end{aligned}
\label{eq:subset_function}
\end{equation}
Here $h_S$ is a subset-specific state within the shared network, and $r_S$ maps it to a task score. The lifting comprises the entire learned mapping from inputs to subset predictions. Collecting these values into a vector $\mathbf f_\theta(x)$ simply supplies coordinates for the resulting function. Each value depends only on its assigned inputs, with parameters learned jointly; the empty-set value fixes the interaction reference. A prediction pathway has order $|S|$, giving unimodal, bimodal, and trimodal pathways. For a fixed sample, write $f(S)=f_\theta(S;x)$.

Subset inclusion organizes the inputs, but it does not make their learned estimates interchangeable. In particular, $f(TAV)$ is one estimate based on complete inputs, whereas the collection $f$ retains estimates at every subset. Their relationship is expressed by the M\"obius coordinates \cite{grabisch2000equivalent}:
\begin{equation}
\Delta_U=\sum_{S\subseteq U}(-1)^{|U|-|S|}f(S),\qquad
f(S)=\sum_{U\subseteq S}\Delta_U.
\label{eq:mobius}
\end{equation}
For example, $\Delta_{TA}=f(TA)-f(T)-f(A)$ extracts the pair increment, whereas $f(TA)=\Delta_T+\Delta_A+\Delta_{TA}$ is cumulative. Geometrically, $\Delta_U$ is the mixed finite difference of $f$ on the subset cube for $U$, anchored at $\varnothing$. Alternating contrasts on edges, faces, and the full cube resolve first-, second-, and third-order increments, respectively (Figure~\ref{fig:spectral_mechanism}(a)). The M\"obius interaction order is $|U|$. Grouping increments gives $g_k=\sum_{|U|=k}\Delta_U$ and $f(TAV)=g_1+g_2+g_3$. Complete-input evaluation assigns a unit response to every order; MoIF regulates this response through its prediction operator.

\subsection{Order-Selective Spectral Regularization}
The prediction operator regulates interaction contributions through subset inclusion. For a set function $f$, define $(K_p f)(S)$ as the expected value of $f(R)$ when each element of $S$ is retained in $R$ independently with probability $p$. The subset zeta matrix $Z_{S,U}=\mathbf1[U\subseteq S]$ maps interaction increments to subset values. In these coordinates, the operator is diagonal \cite{puschel2021discrete}:
\begin{equation}
\begin{aligned}
K_p&=Z\,\operatorname{diag}(p^{|U|})\,Z^{-1},\\
\mathcal F(f)&=\frac{(K_{1/2}f)(\mathcal M)}{1-2^{-M}},\qquad
\hat y=(\mathcal F\circ\Phi_\theta)(x).
\end{aligned}
\label{eq:spectral}
\end{equation}
An increment $\Delta_U$ is included exactly when every element of $U$ is retained, with probability $p^{|U|}$. Interaction order therefore indexes the spectral response. MoIF fixes $p=1/2$ and conditions on a nonempty subset through the denominator in Eq.~\eqref{eq:spectral}. This defines $\mathcal F$ as a deterministic prediction operator, with the expectation providing its algebraic characterization.

The same operator admits an exact subset-domain implementation. At $p=1/2$, all $2^M$ subsets have equal probability; since $f(\varnothing)=0$, nonempty conditioning gives
\begin{equation}
\mathcal F(f)=\frac1{2^M-1}\sum_{\varnothing\ne S\subseteq\mathcal M}f(S)
=\sum_{\varnothing\ne U\subseteq\mathcal M}c_{|U|}\Delta_U,
\qquad c_k=\frac{2^{M-k}}{2^M-1}.
\label{eq:filter}
\end{equation}
The second equality also follows by counting supersets: each $\Delta_U$ occurs in $2^{M-|U|}$ terms of the first sum. For three modalities, singleton, pair, and triplet increments occur in four, two, and one subset values, respectively. Equal coefficients on cumulative predictions therefore induce unequal coefficients on their interaction increments, as shown in Figure~\ref{fig:spectral_mechanism}(b). The resulting prediction is
\begin{equation}
\hat y=\tfrac47g_1+\tfrac27g_2+\tfrac17g_3,
\qquad f(TAV)=g_1+g_2+g_3.
\label{eq:three_orders}
\end{equation}

We term this induced constraint \emph{order-selective spectral regularization}: the response halves with each increase in interaction order, while all orders remain active. Figure~\ref{fig:overview}(b) visualizes its geometry: $D=\operatorname{diag}(4,2,1)/7$ maps the unit sphere in order-sum coordinates to an ellipsoid. The contraction acts on the decision response, leaving higher-order feature construction intact. The first sum in Eq.~\eqref{eq:filter} implements $\mathcal F$ exactly. The task objective trains the lifting through $\mathcal F(f_\theta)$, coupling the learned subset function to this spectral constraint throughout optimization.

The resulting stability depends on the joint variability of the learned increments. With $g=(g_1,g_2,g_3)^\top$, $c=(4,2,1)^\top/7$, and $\Sigma_g$ the covariance across training randomness or repeated perturbations of a fixed example,
\begin{equation}
\operatorname{Var}(\hat y)=c^\top\Sigma_g c.
\label{eq:covariance}
\end{equation}
The full covariance retains dependencies between the nonorthogonal interaction coordinates. This connects the filter response to estimation stability while allowing interactions to reinforce or cancel one another. Section~\ref{sec:spectral_evidence} measures both prediction variability and label-aligned signal, evaluating whether the induced constraint improves their balance.

\subsection{Neural Parameterization of the Learned Lifting}
The lifting $\Phi_\theta$ is parameterized by shared private, pairwise, and trimodal states. Sharing lets the same evidence support several subset estimates, while a distinct scalar projection at each subset preserves its direct predictive role. The hierarchy learns a structured function, rather than seven independent models.

Private encoders retain temporal evidence before any cross-modal interaction. Each stream is projected into a common hidden dimension and encoded independently:
\begin{equation}
\begin{aligned}
H_m&=E_m(X_m,b_m),\qquad h_m=\operatorname{Pool}(H_m,b_m),\\
\operatorname{Pool}(H,b)&=\frac{\sum_\ell b_\ell H_\ell}{\max(1,\sum_\ell b_\ell)}.
\end{aligned}
\label{eq:private}
\end{equation}
Text uses a fine-tuned language model followed by a projection, positional encoding, and a Transformer layer. Audio and vision use separate projections and bidirectional recurrent encoders. Each projection applies a linear map, layer normalization \cite{ba2016layer}, GELU, and dropout. The temporal encoding is added to the projected stream, normalized, and masked. The sequence $H_m$ supplies cross-modal interactions, while its pooled state $h_m$ parameterizes the singleton value $f(\{m\})$.

Balanced reciprocal interaction constructs pair states from both temporal query contexts. For each pair $(m,n)$, pair-specific normalization produces $\bar H_m$ and $\bar H_n$, followed by two separately parameterized attention directions:
\begin{equation}
\begin{aligned}
U_{m\leftarrow n}&=\operatorname{MHA}_{m\leftarrow n}(\bar H_m,\bar H_n,\bar H_n;b_n),\\
U_{n\leftarrow m}&=\operatorname{MHA}_{n\leftarrow m}(\bar H_n,\bar H_m,\bar H_m;b_m).
\end{aligned}
\label{eq:directions}
\end{equation}
The arguments specify queries, keys, values, and the key-validity mask. Pooling each output with its query mask reconciles unequal sequence lengths; equal directional weighting then forms
\begin{equation}
q_{mn}=\tfrac12\left[\operatorname{Pool}(U_{m\leftarrow n},b_m)
+\operatorname{Pool}(U_{n\leftarrow m},b_n)\right].
\label{eq:direction_balance}
\end{equation}
A private residual retains the pooled unimodal evidence alongside this attention update:
\begin{equation}
\begin{aligned}
z_{mn}&=\operatorname{LN}\left(\tfrac12(h_m+h_n)+\operatorname{Drop}(q_{mn})\right),\\
h_{mn}&=z_{mn}+\operatorname{Drop}\left(\operatorname{FFN}(\operatorname{LN}(z_{mn}))\right).
\end{aligned}
\label{eq:pair_state}
\end{equation}
The FFN expands the hidden dimension by a factor of two and projects back after GELU and dropout. Separate modules produce $h_{TA}$, $h_{TV}$, and $h_{AV}$. The attention maps remain sample-dependent, while the equal directional coefficients fix how the two query contexts are combined.

Trimodal composition uses the pair states, preserving their feature content for complete-input prediction. Their concatenation passes through a linear map, normalization, GELU, and dropout to produce $h_{TAV}$. Each subset state is then mapped to its corresponding function value:
\begin{equation}
f_\theta(S;x)=w_S^\top h_S(x_S;\theta)+a_S,
\qquad \hat y=\mathcal F(f_\theta(\cdot;x)).
\label{eq:prediction}
\end{equation}
Here $w_S$ and $a_S$ are subset-specific parameters. These scalar projections complete $\Phi_\theta(x)$; their feature states remain available for higher-order composition. Thus the learned lifting constructs the subset function, and $\mathcal F$ controls its interaction-order response in the decision.

\subsection{Joint Optimization and Subset-Restricted Prediction}
Training couples the learned lifting to its filtered decision. For a batch of $B$ examples, write $f_i=\Phi_\theta(x_i)$ and minimize
\begin{equation}
\mathcal L=\frac1B\sum_i|\mathcal F(f_i)-y_i|
+\frac\lambda{7B}\sum_i\sum_{\varnothing\ne S\subseteq\mathcal M}|f_i(S)-y_i|,
\qquad\lambda=0.1.
\label{eq:objective}
\end{equation}
The primary term optimizes the decision after order-selective filtering. The auxiliary term keeps each subset value task-directed through direct supervision \cite{lee2015deeply}, rather than allowing intermediate predictions to serve only as unconstrained components of the final score. Gradients from both terms update the shared encoders, reciprocal interactions, and subset projections. The fixed operator is thus part of the training computation as well as inference, not a post-training recalibration.

The same construction supports prediction on a restricted input set. At fixed parameters, $x_S=x'_S$ implies $f_\theta(S;x)=f_\theta(S;x')$. For any supplied nonempty set $\mathcal A\subseteq\mathcal M$, restrict the function to its sublattice and evaluate
\begin{equation}
\hat y_{\mathcal A}=\frac1{2^{|\mathcal A|}-1}
\sum_{\varnothing\ne S\subseteq\mathcal A}f_\theta(S;x).
\label{eq:availability}
\end{equation}
The order response becomes $2^{|\mathcal A|-k}/(2^{|\mathcal A|}-1)$ for increments contained in $\mathcal A$. Text and audio, for example, activate T, A, and TA; text alone activates T. This restriction supports the fixed-checkpoint modality-contribution analysis, while the main comparisons use the complete three-modality lattice.

\section{Experiments}
\label{sec:experiments}
\subsection{Tasks and Evaluation}
UR-FUNNY supplies the primary humor-detection evaluation and the controlled mechanism study \cite{hasan2019urfunny}. CMU-MOSI and CMU-MOSEI test cross-task applicability to sentiment prediction, training a separate model on each dataset \cite{zadeh2016mosi,zadeh2018mosei}. Their processed training, validation, and test sizes are (7,614, 980, 994), (1,284, 229, 686), and (16,326, 1,871, 4,659), respectively. All main comparisons use complete text, audio, and vision, retaining the unaligned acoustic and visual features and tokenizing the corresponding utterances. Humor targets are signed binary labels; sentiment targets lie in $[-3,3]$.

Accuracy (Acc-2) and weighted F1 (F1$_w$) assess humor detection, with class-wise precision, recall, and F1 distinguishing humorous and non-humorous behavior. MAE and Pearson correlation (Corr) characterize the continuous score. The sentiment benchmarks additionally use seven-class accuracy (Acc-7), obtained by clipping to $[-3,3]$ and rounding. MoIF's binary metrics threshold predictions at zero and include neutral targets in the positive class: 30 test examples on MOSI and 1,025 on MOSEI. Classification metrics are percentages. Continuous metrics use raw predictions.

\subsection{Implementation and Comparison Protocol}
The common hidden dimension is 128. Text uses fine-tuned RoBERTa-large \cite{liu2019roberta} with at most 64 tokens, followed by a one-layer Transformer with four attention heads and FFN dimension 256. Audio and vision use bidirectional LSTMs \cite{hochreiter1997long} with 64 units per direction. Pair attention has eight heads and FFN dimension 256. Dropout is 0.1 \cite{srivastava2014dropout}; the trimodal map projects 384 concatenated features to 128. The architecture is unchanged across datasets apart from input dimensions.

AdamW \cite{loshchilov2019decoupled} uses learning rate $10^{-5}$, weight decay $10^{-4}$, batch size 32, and gradient clipping at 1. A cosine schedule decreases the rate to $10^{-7}$ over at most 40 epochs. Early stopping has patience 20 after at least 20 epochs; checkpoint selection uses validation MAE. The pretrained text encoder uses mixed precision with dynamic loss scaling, and fusion uses float32. Training and baseline completion used NVIDIA V100 and RTX 5090 GPUs; the controlled-noise analysis used V100 GPUs. Training follows Eq.~\eqref{eq:objective} without additional feature normalization, modality dropout, or input corruption.

The main tables compare thirteen reproduced methods: TFN, LMF, MulT, MISA, Self-MM, ALMT, EMT, LNLN, MPLMM, CASP, DLF, GCE, and TLRRB. TLRRB retains its public model's BERT-base-uncased encoder while sharing the raw utterances, acoustic and visual inputs, splits, validation-MAE selection, and metric functions, including the neutral-sample rule. Results are arithmetic mean$\pm$sample SD over five training seeds per method and dataset. Bold and underline identify the best and second-best displayed means, including ties. Fixed-checkpoint analyses reuse the same five MoIF models per dataset; they do not retrain the constituent subset predictors.

\subsection{Humor Detection and Sentiment Prediction}
Tables~\ref{tab:hmsc_urfunny_main}--\ref{tab:hmsc_mosei_main} report the complete-input comparison. On UR-FUNNY, MoIF achieves the highest accuracy and weighted F1 among the reproduced methods, together with the lowest MAE. Its detection advantage also holds against the recent TLRRB region-modeling approach. MISA has the highest correlation, illustrating that correlation with a continuous score and thresholded humor decisions measure distinct aspects of prediction.

\input{haicme/table2_current}

The sentiment evaluations examine whether the same interaction organization is useful beyond humor detection. On MOSI, MoIF leads all five displayed metrics. On MOSEI, it retains the lowest MAE and highest correlation and Acc-7; GCE has the highest binary accuracy, and its displayed weighted F1 ties MoIF. The cross-task advantage is strongest for continuous and fine-grained prediction, extending the utility of the interaction filter from binary humor recognition to affective intensity estimation.

\subsection{Class-wise Humor Recognition}
Table~\ref{tab:hmsc_urfunny_detail} resolves UR-FUNNY's 504 non-humorous and 490 humorous test examples. MoIF leads both class-wise F1 columns. Its recognition advantage is therefore distributed across the two classes. Compared with GCE, the model trades some humorous recall for higher humorous precision, producing the stronger F1. This balance matters for humor detection: responding to salient expressive cues should not systematically turn non-humorous utterances into false positives.

\input{haicme/urfunny_current}
\clearpage
\input{haicme/moif_analysis}
\FloatBarrier

\section{Discussion}
\label{sec:discussion}
\subsection{Interaction Order as an Estimation Constraint}
Representational capacity and decision response need not increase together. MoIF retains the trimodal state and its supporting pairwise interactions, but does not require every interaction order to enter the estimate with unit response. Lower-order predictions provide the coordinates needed to resolve and regulate this spectrum. Their role extends beyond supplying additional votes: together with the complete-input prediction, they let the shared model implement the order-selective operator.

The UR-FUNNY evidence connects this design to estimation reliability. The declining order-wise SNR shows why increasingly comprehensive interactions need not warrant increasingly strong decision responses. MoIF's geometric response reduces the relative influence of the less reliable orders while retaining their signed contributions. Its benefit is determined jointly by label alignment and cross-order covariance, as verified by reconstructing the measured output SNR. Feature composition remains expressive, while the task loss optimizes the prediction under this response. The result is a learned subset function whose interaction structure supports both discrimination and stable estimation.

\subsection{Scope and Future Directions}
The current model exhaustively represents the seven nonempty subsets of three modalities. Its filter response is fixed by this construction, and the M\"obius coordinates describe learned predictors rather than orthogonal semantic factors. The controlled-noise analysis probes private-feature perturbations on UR-FUNNY; real recording artifacts and distribution changes provide complementary settings for future evaluation. The sentiment results already show that the same construction supports another prediction task, while the family comparisons reveal that the best composition can depend on the metric and dataset.

Larger modality sets make exhaustive subset construction costly. Structured subset sampling and order-dependent responses learned from validation evidence offer routes to scaling the framework. Another direction is to combine quality estimates with the explicit dependency graph, allowing the model to regulate interaction contributions when inputs are unreliable. Both extensions should preserve the connection between an executable prediction rule and its interaction-domain action, so that added flexibility remains interpretable and testable.

\section{Conclusion}
MoIF connects multimodal humor detection to order-selective spectral regularization on a learned subset function. Its central insight is that retaining expressive interactions and regulating their decision contributions are compatible objectives. The M\"obius formulation exposes this distinction, and the measured interaction spectra, signal preservation, and cross-seed stability connect it to improved estimation. Future work will extend the framework to larger modality sets and quality-dependent responses, while preserving the explicit relationship between the prediction rule and its interaction-domain action.

\section*{AI-assisted tools}
The authors used Codex (OpenAI) to assist with manuscript preparation, analysis, visualization, and typesetting. The authors reviewed and verified the outputs and take full responsibility for the content.

\bibliographystyle{IEEEtran}
\bibliography{haicme/current_references}
\end{document}

%% file: arxiv_format.tex
\usepackage{caption}
\makeatletter
\let\ps@pprintTitle\ps@plain

\AtBeginDocument{\typeout{ARXIV-FORMAT: textwidth=\the\textwidth;
  textheight=\the\textheight; body=\f@size pt; baseline=\the\baselineskip}}
\makeatother
\newcommand{\PRmethod}[1]{\parbox[c]{30mm}{\raggedright #1}}
\newcommand{\PRmeanSD}[2]{$#1\pm#2$}

%% file: haicme/table2_current.tex
\begin{table}[p]
\centering
\normalsize
\caption{Complete-input comparison on UR-FUNNY (mean$\pm$SD; five seeds). Classification scores are percentages. Bold and underline indicate the best and second-best means.}
\label{tab:hmsc_urfunny_main}
\setlength{\tabcolsep}{2pt}
\renewcommand{\arraystretch}{1.15}
\begin{tabular*}{\linewidth}{@{\extracolsep{\fill}}lcccc@{}}
\toprule
Method & MAE$\downarrow$ & Corr$\uparrow$ & Acc-2$\uparrow$ & F1$_w$$\uparrow$ \\
\midrule
\PRmethod{TFN~\citep{zadeh2017tensor}~'17} & 0.751$\pm$0.022 & 0.267$\pm$0.023 & 62.7$\pm$1.0 & 62.5$\pm$0.9 \\
\addlinespace[3pt]
\PRmethod{LMF~\citep{liu2018efficient}~'18} & 0.928$\pm$0.005 & 0.143$\pm$0.004 & 55.2$\pm$0.4 & 55.1$\pm$0.5 \\
\addlinespace[3pt]
\PRmethod{MulT~\citep{tsai2019multimodal}~'19} & 0.714$\pm$0.008 & 0.301$\pm$0.006 & 64.7$\pm$0.3 & 64.5$\pm$0.4 \\
\addlinespace[3pt]
\PRmethod{MISA~\citep{hazarika2020misa}~'20} & 0.824$\pm$0.009 & \textbf{0.419}$\pm$0.010 & 66.9$\pm$0.5 & 66.7$\pm$0.5 \\
\addlinespace[3pt]
\PRmethod{Self-MM~\citep{yu2021learning}~'21} & 0.792$\pm$0.016 & 0.222$\pm$0.022 & 60.6$\pm$0.9 & 60.0$\pm$0.9 \\
\addlinespace[3pt]
\PRmethod{ALMT~\citep{zhang2023almt}~'23} & 0.816$\pm$0.007 & 0.201$\pm$0.006 & 59.6$\pm$0.3 & 59.2$\pm$0.1 \\
\addlinespace[3pt]
\PRmethod{EMT~\citep{sun2023emtdlfr}~'24} & 0.774$\pm$0.010 & 0.292$\pm$0.005 & 62.7$\pm$0.3 & 61.9$\pm$1.0 \\
\addlinespace[3pt]
\PRmethod{LNLN~\citep{zhang2024lnln}~'24} & 0.779$\pm$0.008 & 0.290$\pm$0.012 & 63.3$\pm$0.2 & 62.7$\pm$0.7 \\
\addlinespace[3pt]
\PRmethod{MPLMM~\citep{guo2024mplmm}~'24} & 0.717$\pm$0.012 & 0.304$\pm$0.007 & 64.5$\pm$0.5 & 64.0$\pm$0.5 \\
\addlinespace[3pt]
\PRmethod{CASP~\citep{li2025casp}~'25} & 0.755$\pm$0.010 & 0.268$\pm$0.014 & 63.3$\pm$0.6 & 63.2$\pm$0.6 \\
\addlinespace[3pt]
\PRmethod{DLF~\citep{zeng2025dlf}~'25} & 0.814$\pm$0.020 & 0.257$\pm$0.033 & 61.0$\pm$2.1 & 60.6$\pm$2.0 \\
\addlinespace[3pt]
\PRmethod{GCE~\citep{huang2026grading}~'26} & \underline{0.655}$\pm$0.014 & 0.363$\pm$0.015 & 67.2$\pm$0.8 & 67.0$\pm$0.6 \\
\addlinespace[3pt]
\PRmethod{TLRRB~\citep{li2026tlrrb}~'26} & 0.656$\pm$0.018 & 0.355$\pm$0.017 & \underline{67.3}$\pm$0.9 & \underline{67.2}$\pm$0.9 \\
\addlinespace[3pt]
\midrule
\PRmethod{\textbf{MoIF (Ours)}} & \textbf{0.599}$\pm$0.008 & \underline{0.417}$\pm$0.007 & \textbf{70.6}$\pm$0.4 & \textbf{70.6}$\pm$0.4 \\
\bottomrule
\end{tabular*}
\end{table}

\begin{table}[p]
\centering
\normalsize
\caption{Complete-input comparison on CMU-MOSI (mean$\pm$SD; five seeds). Classification scores are percentages. Bold and underline indicate the best and second-best means.}
\label{tab:hmsc_mosi_main}
\setlength{\tabcolsep}{2pt}
\renewcommand{\arraystretch}{1.15}
\begin{tabular*}{\linewidth}{@{\extracolsep{\fill}}lccccc@{}}
\toprule
Method & MAE$\downarrow$ & Corr$\uparrow$ & Acc-7$\uparrow$ & Acc-2$\uparrow$ & F1$_w$$\uparrow$ \\
\midrule
\PRmethod{TFN~\citep{zadeh2017tensor}~'17} & 0.903$\pm$0.011 & 0.659$\pm$0.008 & 36.6$\pm$0.03 & 78.5$\pm$0.4 & 78.5$\pm$0.4 \\
\addlinespace[3pt]
\PRmethod{LMF~\citep{liu2018efficient}~'18} & 0.941$\pm$0.008 & 0.656$\pm$0.005 & 34.3$\pm$0.3 & 76.5$\pm$0.3 & 76.7$\pm$0.3 \\
\addlinespace[3pt]
\PRmethod{MulT~\citep{tsai2019multimodal}~'19} & 0.898$\pm$0.009 & 0.682$\pm$0.002 & 37.8$\pm$0.0 & 78.9$\pm$0.6 & 78.2$\pm$0.6 \\
\addlinespace[3pt]
\PRmethod{MISA~\citep{hazarika2020misa}~'20} & 0.940$\pm$0.012 & 0.668$\pm$0.009 & 35.0$\pm$1.3 & 77.8$\pm$0.8 & 77.5$\pm$0.8 \\
\addlinespace[3pt]
\PRmethod{Self-MM~\citep{yu2021learning}~'21} & 0.788$\pm$0.012 & 0.742$\pm$0.010 & 43.3$\pm$1.2 & 80.8$\pm$0.5 & 80.1$\pm$0.4 \\
\addlinespace[3pt]
\PRmethod{ALMT~\citep{zhang2023almt}~'23} & 0.749$\pm$0.007 & 0.772$\pm$0.005 & 44.1$\pm$0.9 & 81.0$\pm$1.1 & 80.5$\pm$1.0 \\
\addlinespace[3pt]
\PRmethod{EMT~\citep{sun2023emtdlfr}~'24} & \underline{0.715}$\pm$0.009 & \underline{0.794}$\pm$0.007 & \underline{47.4}$\pm$1.4 & 82.2$\pm$0.2 & 82.1$\pm$0.2 \\
\addlinespace[3pt]
\PRmethod{LNLN~\citep{zhang2024lnln}~'24} & 0.753$\pm$0.009 & 0.760$\pm$0.001 & 44.8$\pm$2.0 & 80.9$\pm$0.4 & 80.3$\pm$0.4 \\
\addlinespace[3pt]
\PRmethod{MPLMM~\citep{guo2024mplmm}~'24} & 0.902$\pm$0.007 & 0.679$\pm$0.003 & 37.2$\pm$0.8 & 78.9$\pm$0.4 & 78.2$\pm$0.4 \\
\addlinespace[3pt]
\PRmethod{CASP~\citep{li2025casp}~'25} & 0.876$\pm$0.006 & 0.693$\pm$0.003 & 36.3$\pm$0.6 & 79.4$\pm$0.5 & 79.1$\pm$0.5 \\
\addlinespace[3pt]
\PRmethod{DLF~\citep{zeng2025dlf}~'25} & 0.794$\pm$0.013 & 0.770$\pm$0.011 & 41.2$\pm$1.0 & 81.1$\pm$0.3 & 82.0$\pm$0.3 \\
\addlinespace[3pt]
\PRmethod{GCE~\citep{huang2026grading}~'26} & 0.748$\pm$0.012 & 0.781$\pm$0.009 & 43.6$\pm$0.6 & 82.5$\pm$1.2 & 82.5$\pm$1.2 \\
\addlinespace[3pt]
\PRmethod{TLRRB~\citep{li2026tlrrb}~'26} & 0.731$\pm$0.018 & 0.790$\pm$0.012 & 45.8$\pm$1.4 & \underline{82.8}$\pm$0.7 & \underline{82.8}$\pm$0.7 \\
\addlinespace[3pt]
\midrule
\PRmethod{\textbf{MoIF (Ours)}} & \textbf{0.645}$\pm$0.017 & \textbf{0.845}$\pm$0.008 & \textbf{47.7}$\pm$1.9 & \textbf{84.9}$\pm$0.4 & \textbf{84.9}$\pm$0.4 \\
\bottomrule
\end{tabular*}
\end{table}

\begin{table}[p]
\centering
\normalsize
\caption{Complete-input comparison on CMU-MOSEI (mean$\pm$SD; five seeds). Classification scores are percentages. Bold and underline indicate the best and second-best means.}
\label{tab:hmsc_mosei_main}
\setlength{\tabcolsep}{2pt}
\renewcommand{\arraystretch}{1.15}
\begin{tabular*}{\linewidth}{@{\extracolsep{\fill}}lccccc@{}}
\toprule
Method & MAE$\downarrow$ & Corr$\uparrow$ & Acc-7$\uparrow$ & Acc-2$\uparrow$ & F1$_w$$\uparrow$ \\
\midrule
\PRmethod{TFN~\citep{zadeh2017tensor}~'17} & 0.577$\pm$0.008 & 0.711$\pm$0.007 & 50.7$\pm$0.3 & 79.0$\pm$1.3 & 79.2$\pm$1.4 \\
\addlinespace[3pt]
\PRmethod{LMF~\citep{liu2018efficient}~'18} & 0.584$\pm$0.004 & 0.714$\pm$0.008 & 50.9$\pm$0.5 & 78.7$\pm$2.3 & 79.3$\pm$2.0 \\
\addlinespace[3pt]
\PRmethod{MulT~\citep{tsai2019multimodal}~'19} & 0.551$\pm$0.002 & 0.747$\pm$0.002 & 53.7$\pm$0.8 & 77.4$\pm$0.3 & 79.1$\pm$0.3 \\
\addlinespace[3pt]
\PRmethod{MISA~\citep{hazarika2020misa}~'20} & 0.580$\pm$0.004 & 0.719$\pm$0.003 & 50.6$\pm$0.1 & 80.6$\pm$0.2 & 81.0$\pm$0.2 \\
\addlinespace[3pt]
\PRmethod{Self-MM~\citep{yu2021learning}~'21} & 0.588$\pm$0.006 & 0.738$\pm$0.007 & 52.3$\pm$0.4 & \underline{82.5}$\pm$1.1 & \underline{82.2}$\pm$1.0 \\
\addlinespace[3pt]
\PRmethod{ALMT~\citep{zhang2023almt}~'23} & \underline{0.530}$\pm$0.004 & \underline{0.776}$\pm$0.002 & 53.2$\pm$0.8 & 81.3$\pm$0.6 & 82.0$\pm$0.7 \\
\addlinespace[3pt]
\PRmethod{EMT~\citep{sun2023emtdlfr}~'24} & 0.531$\pm$0.004 & 0.762$\pm$0.003 & 52.9$\pm$0.7 & 82.2$\pm$1.8 & 81.9$\pm$1.2 \\
\addlinespace[3pt]
\PRmethod{LNLN~\citep{zhang2024lnln}~'24} & 0.550$\pm$0.001 & 0.748$\pm$0.001 & 53.1$\pm$0.4 & 79.1$\pm$0.5 & 80.3$\pm$0.7 \\
\addlinespace[3pt]
\PRmethod{MPLMM~\citep{guo2024mplmm}~'24} & 0.549$\pm$0.003 & 0.747$\pm$0.003 & 53.5$\pm$0.1 & 80.1$\pm$1.7 & 81.8$\pm$0.9 \\
\addlinespace[3pt]
\PRmethod{CASP~\citep{li2025casp}~'25} & 0.550$\pm$0.004 & 0.742$\pm$0.003 & 53.1$\pm$0.3 & 80.5$\pm$1.2 & 81.1$\pm$1.0 \\
\addlinespace[3pt]
\PRmethod{DLF~\citep{zeng2025dlf}~'25} & 0.555$\pm$0.006 & 0.735$\pm$0.003 & 52.6$\pm$0.7 & 81.1$\pm$1.2 & 80.8$\pm$0.8 \\
\addlinespace[3pt]
\PRmethod{GCE~\citep{huang2026grading}~'26} & 0.558$\pm$0.004 & 0.757$\pm$0.002 & 52.4$\pm$0.6 & \textbf{83.4}$\pm$1.3 & \textbf{83.0}$\pm$1.1 \\
\addlinespace[3pt]
\PRmethod{TLRRB~\citep{li2026tlrrb}~'26} & 0.535$\pm$0.010 & 0.770$\pm$0.011 & \underline{53.8}$\pm$0.9 & 81.6$\pm$1.3 & 82.1$\pm$1.1 \\
\addlinespace[3pt]
\midrule
\PRmethod{\textbf{MoIF (Ours)}} & \textbf{0.492}$\pm$0.004 & \textbf{0.810}$\pm$0.003 & \textbf{55.8}$\pm$0.3 & \underline{82.5}$\pm$2.0 & \textbf{83.0}$\pm$1.8 \\
\bottomrule
\end{tabular*}
\end{table}

%% file: haicme/urfunny_current.tex
\begin{table}[p]
\centering
\normalsize
\caption{UR-FUNNY class-wise comparison (mean$\pm$SD; five seeds). P and R denote precision and recall. All scores are percentages. Best and second-best means are bold and underlined.}
\label{tab:hmsc_urfunny_detail}
\setlength{\tabcolsep}{2pt}
\renewcommand{\arraystretch}{1.15}
\begin{tabular*}{\linewidth}{@{\extracolsep{\fill}}lcccccc@{}}
\toprule
& \multicolumn{3}{c}{Non-humorous} & \multicolumn{3}{c}{Humorous} \\
\cmidrule(lr){2-4}\cmidrule(l){5-7}
Method & P & R & F1 & P & R & F1 \\
\midrule
\PRmethod{TFN~\citep{zadeh2017tensor}~'17} & 64.4$\pm$2.9 & 60.4$\pm$6.3 & 62.0$\pm$2.1 & 61.7$\pm$1.4 & 65.0$\pm$7.7 & 63.0$\pm$3.2 \\
\addlinespace[3pt]
\PRmethod{LMF~\citep{liu2018efficient}~'18} & 56.4$\pm$0.3 & 50.9$\pm$2.0 & 53.5$\pm$1.2 & 54.1$\pm$0.5 & 59.6$\pm$1.2 & 56.7$\pm$0.3 \\
\addlinespace[3pt]
\PRmethod{MulT~\citep{tsai2019multimodal}~'19} & 67.8$\pm$0.1 & 57.8$\pm$1.3 & 62.4$\pm$0.7 & 62.3$\pm$0.5 & 71.7$\pm$0.7 & 66.7$\pm$0.0 \\
\addlinespace[3pt]
\PRmethod{MISA~\citep{hazarika2020misa}~'20} & 69.8$\pm$0.5 & 61.0$\pm$0.5 & 65.1$\pm$0.5 & 64.5$\pm$0.4 & \underline{72.9}$\pm$0.4 & 68.4$\pm$0.4 \\
\addlinespace[3pt]
\PRmethod{Self-MM~\citep{yu2021learning}~'21} & 59.1$\pm$0.7 & \textbf{71.9}$\pm$1.3 & 64.9$\pm$0.9 & 62.9$\pm$1.3 & 48.9$\pm$0.5 & 55.0$\pm$0.8 \\
\addlinespace[3pt]
\PRmethod{ALMT~\citep{zhang2023almt}~'23} & 58.6$\pm$0.1 & 69.2$\pm$2.6 & 63.5$\pm$1.0 & 61.2$\pm$1.0 & 49.7$\pm$2.1 & 54.8$\pm$0.9 \\
\addlinespace[3pt]
\PRmethod{EMT~\citep{sun2023emtdlfr}~'24} & 68.9$\pm$3.2 & 49.4$\pm$6.9 & 57.1$\pm$3.6 & 59.6$\pm$1.2 & \textbf{76.4}$\pm$6.6 & 66.8$\pm$1.8 \\
\addlinespace[3pt]
\PRmethod{LNLN~\citep{zhang2024lnln}~'24} & 67.5$\pm$3.9 & 55.7$\pm$10.0 & 60.1$\pm$4.4 & 61.3$\pm$2.1 & 71.2$\pm$10.0 & 65.4$\pm$3.1 \\
\addlinespace[3pt]
\PRmethod{MPLMM~\citep{guo2024mplmm}~'24} & 69.8$\pm$0.4 & 52.9$\pm$0.9 & 60.2$\pm$0.7 & 61.2$\pm$0.5 & \textbf{76.4}$\pm$0.1 & 68.0$\pm$0.3 \\
\addlinespace[3pt]
\PRmethod{CASP~\citep{li2025casp}~'25} & 65.1$\pm$0.7 & 59.5$\pm$0.6 & 62.2$\pm$0.6 & 61.7$\pm$0.6 & 67.1$\pm$0.6 & 64.3$\pm$0.6 \\
\addlinespace[3pt]
\PRmethod{DLF~\citep{zeng2025dlf}~'25} & 63.2$\pm$4.9 & 58.8$\pm$7.6 & 60.3$\pm$1.8 & 59.8$\pm$0.2 & 63.2$\pm$12.1 & 60.9$\pm$5.9 \\
\addlinespace[3pt]
\PRmethod{GCE~\citep{huang2026grading}~'26} & \underline{70.2}$\pm$2.4 & 61.6$\pm$2.7 & 65.5$\pm$0.5 & 64.9$\pm$0.3 & \underline{72.9}$\pm$4.3 & \underline{68.6}$\pm$1.8 \\
\addlinespace[3pt]
\PRmethod{TLRRB~\citep{li2026tlrrb}~'26} & 68.9$\pm$1.5 & 64.8$\pm$4.4 & \underline{66.7}$\pm$1.9 & \underline{65.9}$\pm$1.7 & 69.8$\pm$3.9 & 67.7$\pm$1.3 \\
\addlinespace[3pt]
\midrule
\PRmethod{\textbf{MoIF (Ours)}} & \textbf{71.4}$\pm$0.9 & \underline{70.1}$\pm$2.1 & \textbf{70.7}$\pm$0.8 & \textbf{69.8}$\pm$1.0 & 71.1$\pm$2.1 & \textbf{70.4}$\pm$0.7 \\
\bottomrule
\end{tabular*}
\end{table}

%% file: haicme/moif_analysis.tex
\subsection{Interaction Spectra and Estimation Stability}
\label{sec:spectral_evidence}
\begin{figure*}[p]
\centering
\includegraphics[width=\linewidth,height=0.80\textheight,keepaspectratio]{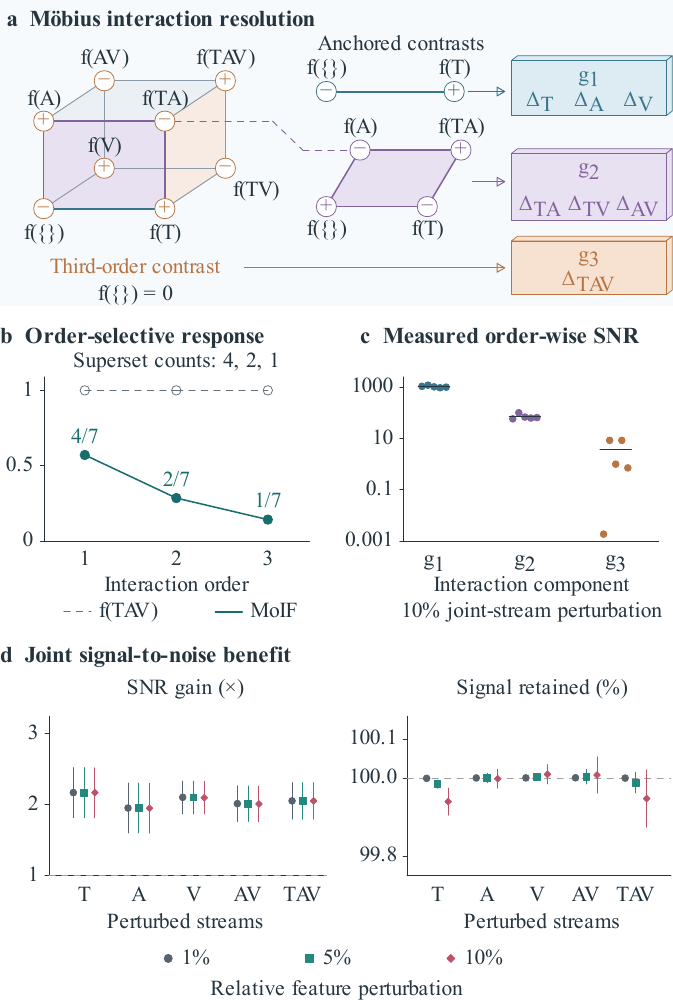}
\caption{Interaction resolution and filtering evidence. (a) Vertex signs define the third-order cube contrast; edge and face expansions illustrate lower-order contrasts, with $f(\varnothing)=0$ and $g_k=\sum_{|U|=k}\Delta_U$. (b) Unit response of $f(TAV)$ versus MoIF's $(4,2,1)/7$. (c) UR-FUNNY order-wise SNR at 10\% joint-stream perturbation: five checkpoints and arithmetic means, on a log scale. (d) SNR gain over $f(TAV)$ and signal retention relative to MoIF's clean output (mean$\pm$SD, five checkpoints). Shapes distinguish perturbation magnitudes. Joint SNR uses signed alignment and full covariance, Eq.~\eqref{eq:joint_snr}. Panels (a,b) are algebraic; (c,d) are measured.}
\label{fig:spectral_mechanism}
\end{figure*}

The mechanism study tests whether the induced order response improves the balance between task-aligned signal and perturbation sensitivity. Figure~\ref{fig:spectral_mechanism} connects the discrete interaction contrasts to three quantities: the operator's response, the reliability of the learned interaction orders, and the resulting prediction SNR. Panels (a,b) describe the operator; panels (c,d) evaluate its action on UR-FUNNY.

With parameters fixed, we perturb valid private-encoder features immediately before balanced reciprocal interaction. Gaussian directions are normalized separately for each sample and modality to relative Frobenius magnitudes of 1\%, 5\%, and 10\%. The affected streams are T, A, V, AV, or TAV, giving 75 seed--condition combinations. Each uses 16 shared perturbation draws for all subset predictors. For a predictor $f$, clean output $f_0$, and perturbed output $f_r$, we measure
\begin{equation}
\begin{aligned}
S(f_0)&=\frac{\operatorname{Cov}_i(f_0,y)^2}{\operatorname{Var}_i(y)},\\
N(f)&=\mathbb E_i\operatorname{Var}_r[f_r(i)],\qquad
\operatorname{SNR}(f)=\frac{S(f_0)}{N(f)}.
\end{aligned}
\label{eq:signal_noise}
\end{equation}
For a binary target, $S$ is the output variance explained by the two class means. Signal retention is $S(\mathbb E_r f_r)/S(f_0)$. Empirical perturbation variances use divisor 16. The SNR is invariant to nonzero global score scaling; component covariances are retained when evaluating Eq.~\eqref{eq:covariance}.

The learned spectrum exhibits a pronounced reliability gradient. At 10\% joint-stream perturbation, the mean SNRs of $g_1$, $g_2$, and $g_3$ are 1,050.7, 71.0, and 3.7. Figure~\ref{fig:spectral_mechanism}(c) shows all five checkpoints; the descending order holds in every tested seed--condition combination. Higher-order increments retain less label-aligned signal relative to their perturbation variability. Complete-input evaluation gives these unequal-reliability components the same unit response, whereas MoIF assigns progressively smaller coefficients. The measured spectrum thus supports the direction of the induced selectivity.

The output benefit depends on how these components combine, not on their separate SNRs alone. For the induced response $\mathbf c=(4,2,1)^\top/7$, let $a_k=\operatorname{Cov}_i(g_{k,0},y)$ and $C_{kl}=\mathbb E_i\operatorname{Cov}_r(g_{k,r},g_{l,r})$. Signed alignment and perturbation covariance determine the joint SNR:
\begin{equation}
\operatorname{SNR}(\hat y)=
\frac{(\mathbf c^\top\mathbf a)^2}
{\operatorname{Var}_i(y)\,\mathbf c^\top C\mathbf c}.
\label{eq:joint_snr}
\end{equation}
The numerator retains signed label alignment; the denominator includes all cross-order perturbation covariances. Replacing $\mathbf c$ by $(1,1,1)^\top$ recovers the co-trained $f(TAV)$ statistics. Both reconstructions match the directly measured SNR in every condition. The improvement therefore reflects the joint spectral response, rather than an average of the three component SNRs.

Figure~\ref{fig:spectral_mechanism}(d) confirms the improvement: MoIF reduces perturbation variance in all 75 seed--condition combinations, with SNR gains of 1.62--2.79 times over $f(TAV)$ and signal retention of 99.849--100.061\% relative to its own clean output. Table~\ref{tab:noise_stability} reports the largest perturbation magnitude; the figure includes both smaller magnitudes. A float32 repeat preserves the ordering in all 75 combinations. These paired V100 evaluations leave the historical main tables unchanged. Aggregate replay MAE differs from the archived outputs by at most 0.000344, with 0--2 changed decisions per 994 examples; replay diagnostics and available configuration records accompany the source package.

\begin{table}[!htbp]
\centering
\caption{UR-FUNNY stability under 10\% relative private-feature perturbations (mean$\pm$SD over five checkpoints). Noise reduction and SNR gain compare MoIF with $f(TAV)$. Signal retention compares MoIF with its own clean output.}
\label{tab:noise_stability}
\setlength{\tabcolsep}{3pt}
\renewcommand{\PRmeanSD}[2]{$#1\pm#2$}
\begin{tabular*}{\linewidth}{@{\extracolsep{\fill}}lccc@{}}
\toprule
Perturbed & Noise reduction & SNR gain & Signal retention\\
streams & (\%) & ($\times$) & (\%)\\
\midrule
T & \PRmeanSD{65.07}{6.54} & \PRmeanSD{2.169}{0.350} & \PRmeanSD{99.939}{0.035}\\
A & \PRmeanSD{61.02}{7.71} & \PRmeanSD{1.950}{0.350} & \PRmeanSD{99.998}{0.024}\\
V & \PRmeanSD{64.14}{6.04} & \PRmeanSD{2.098}{0.229} & \PRmeanSD{100.010}{0.025}\\
AV & \PRmeanSD{62.61}{5.79} & \PRmeanSD{2.010}{0.248} & \PRmeanSD{100.008}{0.047}\\
TAV & \PRmeanSD{63.38}{5.69} & \PRmeanSD{2.052}{0.257} & \PRmeanSD{99.947}{0.073}\\
\bottomrule
\end{tabular*}
\end{table}

Training variability provides a complementary test. We calculate each example's prediction variance across five independently trained checkpoints and average over the test set. Relative to $f(TAV)$, MoIF reduces this variance by 32.77\% on UR-FUNNY, 54.03\% on MOSI, and 62.20\% on MOSEI. Together with the perturbation analysis, this supports a consistent estimation benefit across training realizations and controlled input-feature changes.

\subsection{Prediction Pathways and Individual Interaction Spectra}
\label{sec:expert_analysis}
\begin{figure*}[p]
\centering
\includegraphics[width=\linewidth,height=0.80\textheight,keepaspectratio]{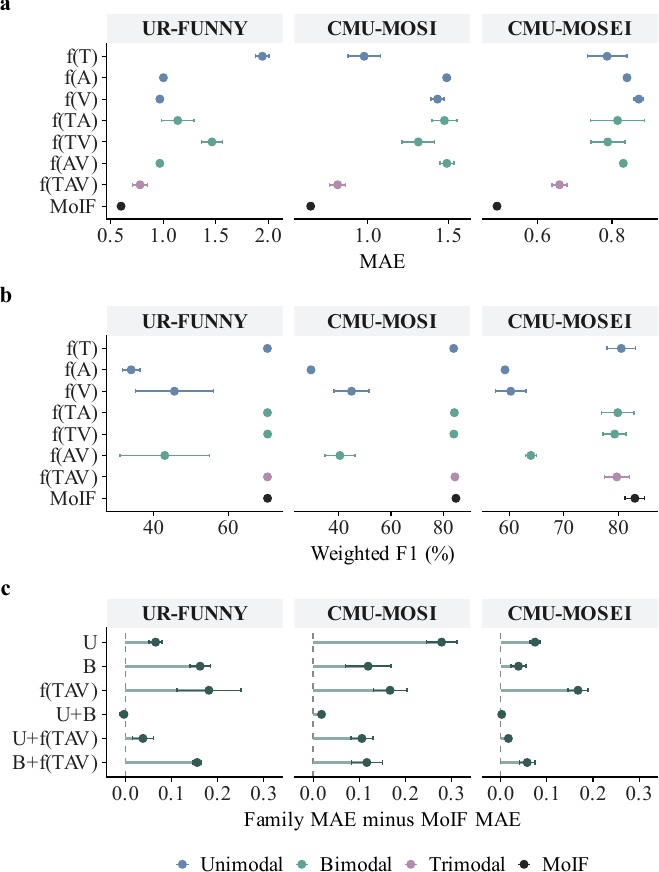}
\caption{Subset predictors and prediction-family aggregates. (a) MAE and (b) weighted F1 of $f(S)$ and MoIF. (c) Family MAE minus MoIF MAE; positive values favor MoIF. U and B contain singleton and pair predictors; $f(TAV)$ is the trimodal predictor. A family combination uniformly averages its included predictors. Points and error bars show mean$\pm$SD over five checkpoints per dataset.}
\label{fig:expert_collaboration}
\end{figure*}

Strong multimodal performance alone does not identify how interactions contribute to prediction \cite{hessel2020interactions}. Figure~\ref{fig:expert_collaboration} compares the filtered estimate with individual subset predictors and prediction-family averages, holding the five trained checkpoints and test examples fixed. The unimodal family U comprises $f(T)$, $f(A)$, and $f(V)$; the bimodal family B comprises $f(TA)$, $f(TV)$, and $f(AV)$. Family labels specify which predictions are averaged, whereas $g_k$ denotes an interaction-order component in M\"obius coordinates.

MoIF has lower MAE than every individual subset predictor and either the unimodal or bimodal family on all three datasets. Complete-input representation learning therefore does not make the lower-order prediction pathways redundant. Their contribution to the filtered estimate changes the interaction response from the full-set predictor's unit coefficients to the geometric coefficients in Eq.~\eqref{eq:three_orders}. Classification and target-magnitude estimation emphasize different aspects: the text predictor nearly matches MoIF's UR-FUNNY weighted F1, and the unimodal family has higher weighted F1 on MOSEI. The most consistent within-model advantage is in continuous estimation; the humor-detection advantage over other methods is established by the main comparison.

The third-order prediction pathway makes a task-dependent incremental contribution. Including $f(TAV)$ in the fixed U+B aggregate improves MAE on both sentiment datasets, whereas U+B has slightly lower MAE on UR-FUNNY. This comparison preserves the contribution analysis at the prediction-pathway level: including the trimodal prediction changes the spectrum entering the aggregate, rather than making the lower-order pathways dispensable.

\begin{figure*}[p]
\centering
\includegraphics[width=\linewidth,height=0.80\textheight,keepaspectratio]{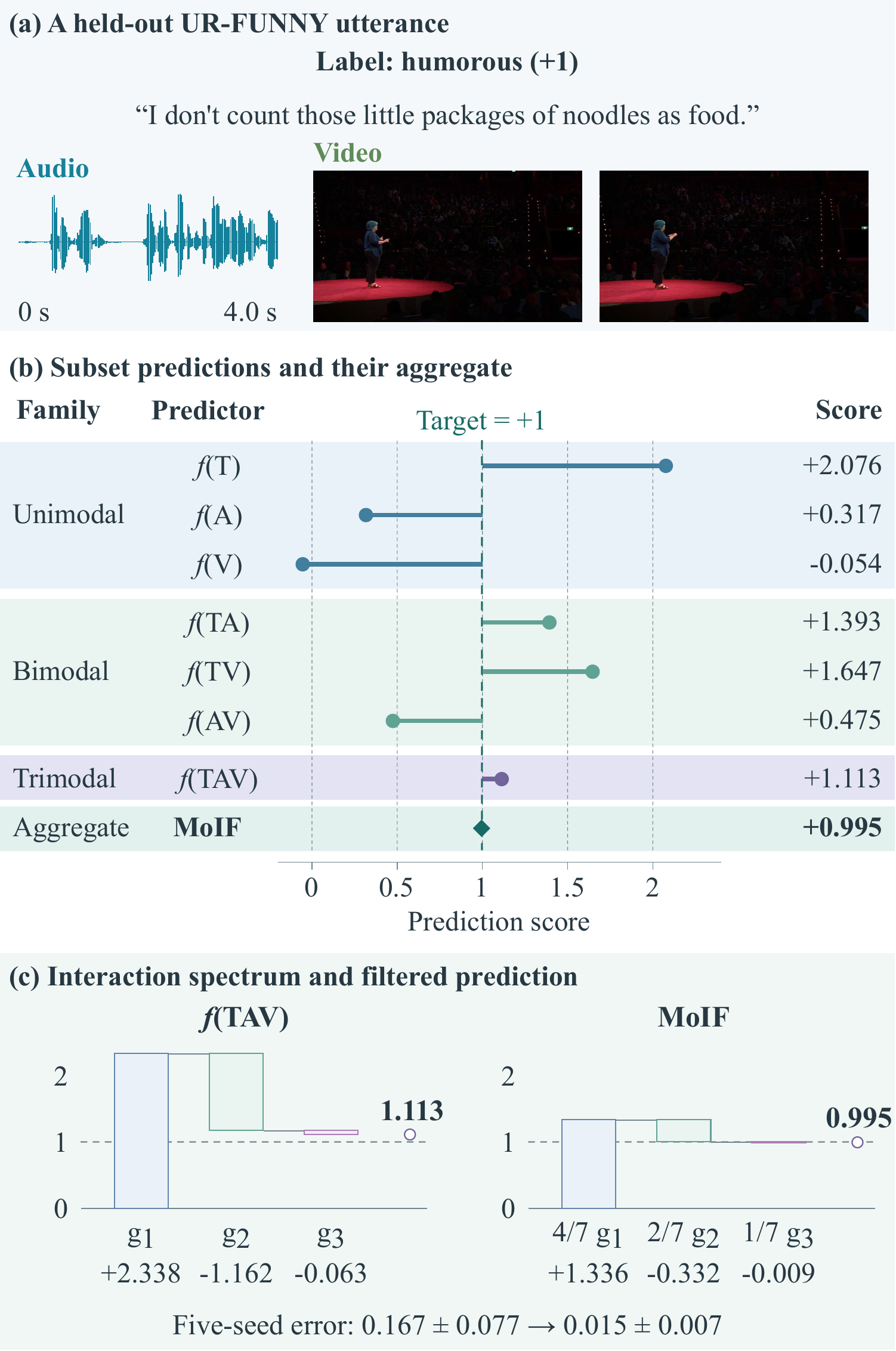}
\caption{An illustrative interaction-filtering case, selected after examining UR-FUNNY test predictions: example 4274 (humorous, target +1). (a) Text and audiovisual excerpt. (b) Seven subset predictions $f(S)$ and MoIF at seed 42. (c) Signed order contributions reconstruct $f(TAV)$ on the left and the filtered score on the right; dashed lines mark the target. The final line reports mean$\pm$SD of absolute error over five seeds, from $f(TAV)$ to MoIF. Contributions are computed before rounding.}
\label{fig:prediction_case}
\end{figure*}

Figure~\ref{fig:prediction_case} resolves the same mechanism for a humorous utterance. Its text-containing predictions exceed the target, while the audio and visual predictions fall below it. The M\"obius decomposition exposes the signed increments underlying these cumulative values. Summing the three order components reconstructs $f(TAV)$; applying MoIF's order response to those same components yields a score closer to the target. Across all five seeds, the joint estimate is closer than every individual subset prediction. Both $f(TAV)$ and MoIF classify this example correctly; the spectral view explains the improvement in its continuous estimate.

\subsection{Input Context and Restricted Interaction Structure}
\label{sec:modality_analysis}
Conditional modality utilization asks what a stream adds to evidence already available \cite{wu2022greedy}. We compare the error of $\hat y_{\mathcal A}$ before and after each of nine additions to a nonempty input set $\mathcal A$, using Eq.~\eqref{eq:availability}. Set labels such as $\{T,A\}$ specify available inputs, not the individual predictor $f(TA)$. The same saved subset predictions are reused; adding an input expands the eligible sublattice and changes its aggregate response.

\begin{figure*}[p]
\centering
\includegraphics[width=\linewidth,height=0.80\textheight,keepaspectratio]{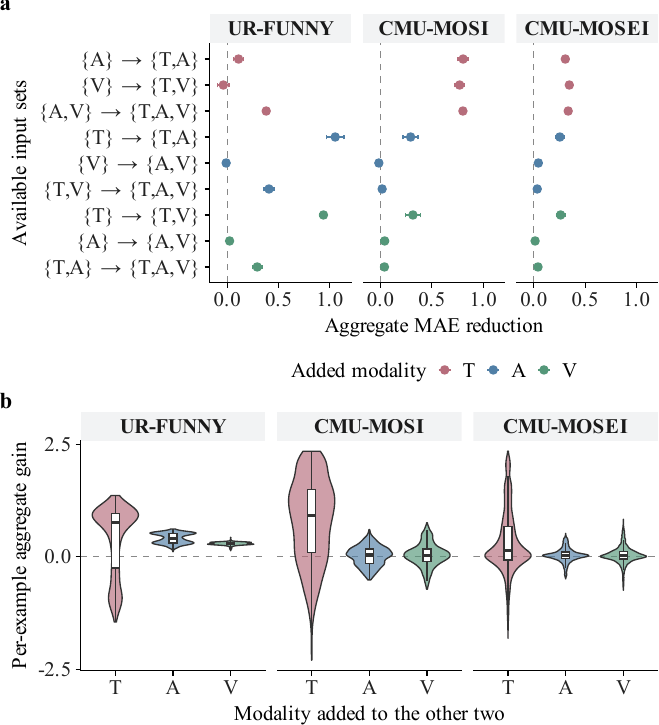}
\caption{Modality additions under subset-restricted prediction. Colors identify the added modality in both panels. (a) Aggregate MAE reductions (mean$\pm$SD over five seeds). (b) Per-example error reductions when adding a stream to the other two, averaged over seeds. Violin widths encode density normalized to equal maximum width, not sample counts. Boxes show medians and interquartile ranges (IQR); whiskers reach observations within 1.5 IQR of the box. Positive values favor inclusion. The aggregate for $\mathcal A=\{T,A,V\}$ uses all seven subset predictions.}
\label{fig:modality_contribution}
\end{figure*}

Figure~\ref{fig:modality_contribution} shows that the resulting benefit depends on the input context. Adding text to $\{A,V\}$ improves mean MAE on every dataset, yet adding it to $\{V\}$ alone can worsen UR-FUNNY performance. Audio and vision each help when added to the other two streams across all three datasets, although positive mean gains coexist with individual increases in error. This complements the spectral analysis: an input's predictive value depends on the interaction structure it enables and on the operator combining that structure, not solely on its standalone strength.

\FloatBarrier
\subsection{Independently Trained Directional Controls}
\label{sec:direction_controls}
Table~\ref{tab:direction_controls} isolates balanced reciprocal interaction. All configurations retain the seven subset predictors, private residuals, auxiliary supervision, and training schedule. The single-direction control disables reverse attention. The learned-weight control replaces equal directional weighting with a sigmoid scalar computed from the concatenated directional summaries. Each configuration is independently trained with the same five UR-FUNNY seeds and selected by validation MAE.

\begin{table}[!htbp]
\centering
\caption{Independently trained directional controls on UR-FUNNY (mean$\pm$SD over five seeds). Classification scores are percentages. Best and second-best means are bold and underlined.}
\label{tab:direction_controls}
\setlength{\tabcolsep}{2pt}
\begin{tabular*}{\linewidth}{@{\extracolsep{\fill}}lcccc@{}}
\toprule
\PRmethod{Reciprocal interaction} & MAE$\downarrow$ & Corr$\uparrow$ & Acc-2$\uparrow$ & F1$_w$$\uparrow$\\
\midrule
\PRmethod{Single direction} & \PRmeanSD{0.617}{0.015} & \PRmeanSD{\underline{0.406}}{0.015} & \PRmeanSD{69.98}{0.75} & \PRmeanSD{69.88}{0.81}\\
\PRmethod{Learned weighting} & \PRmeanSD{\underline{0.612}}{0.016} & \PRmeanSD{0.404}{0.013} & \PRmeanSD{\underline{70.00}}{0.68} & \PRmeanSD{\underline{69.99}}{0.68}\\
\PRmethod{Balanced reciprocal interaction (MoIF)} & \PRmeanSD{\mathbf{0.599}}{0.008} & \PRmeanSD{\mathbf{0.417}}{0.007} & \PRmeanSD{\mathbf{70.58}}{0.44} & \PRmeanSD{\mathbf{70.57}}{0.44}\\
\bottomrule
\end{tabular*}
\end{table}

Balanced reciprocal interaction improves all four mean metrics over both controls. Retaining both query contexts is useful, while their equal combination achieves the strongest result in this comparison. This tests the neural parameterization that learns bimodal predictors; the spectral analysis above characterizes the operator regulating their interaction contributions.